\documentclass[prb,aps,epsf,twocolumn,showpacs,10pt]{revtex4-1}
\usepackage{graphicx,amsfonts,times,bm,amsmath,verbatim,color,array}

\renewcommand{\(}{\left(}
\renewcommand{\)}{\right)}

\begin{document}
\title{Nernst effect in Dirac and inversion asymmetric Weyl semimetals}

\author{Girish Sharma$^{1,2}$}
\author{Christopher Moore$^1$}
\author{Subhodip Saha$^{3}$}
\author{Sumanta Tewari$^1$}

\affiliation{$^1$Department of Physics and Astronomy, Clemson University, Clemson, SC 29634\\
$^2$Department of Physics, Virginia Tech, Blacksburg, VA 24061, U.S.A\\
$^3$ Department of Physics, Indian Institute of Technology, Kharagpur, 721302, India}

\begin{abstract}
Dirac semimetals are three dimensional analog of graphene with massless Dirac fermions as low energy electronic excitations. In contrast to Weyl semimetals, the point nodes in the bulk spectrum of topological Dirac semimetals have a vanishing Chern number, but can yet be stable due to the existence of crystalline symmetries such as uniaxial (discrete) rotation symmetry. We consider a model low-energy Hamiltonian appropriate for the recently discovered topological Dirac semimetal Cd$_3$As$_2$, and calculate the Nernst response within semiclassical Boltzmann dynamics in the relaxation time approximation. We show that, for small chemical potentials near the Dirac points, the low temperature, low magnetic field, Nernst response is dominated by \textit{anomalous} Nernst effect, arising from a non-trivial profile of Berry curvature on the Fermi surface. Although the Nernst coefficient (both anomalous as well as conventional) vanish in the limit of zero magnetic field, the low temperature, low magnetic field, Nernst response, which has an almost step like profile near $\mathbf{B}=0$, serves as an effective experimental probe of anomalous Nernst effect in topological Dirac semimetals protected by crystalline symmetries. Additionally, we also calculate the Nernst response for a lattice model of an inversion asymmetric Weyl semimetal for which, in contrast to the case of Dirac semimetal, we find that the conventional Nernst response dominates over the anomalous. Our calculations in this paper on Nernst response of Dirac semimetal and inversion broken Weyl semimetal are directly relevant to recent experiments on Cd$_3$As$_2$ (Dirac semimetal) and NbP (inversion broken Weyl semimetal) respectively.
\end{abstract}

\maketitle

\section{Introduction}
Dirac semimetals (DSM) are three-dimensional (3D) analog of graphene, with point nodes in the bulk energy spectrum supporting low energy excitations with relativistic energy momentum relations resembling massless Dirac fermions~\cite{Murakami1:2007, Murakami2:2007}. In principle they can arise at the quantum critical point between a 3D  topological insulator and a conventional insulator with fine tuning of an external parameter. Topological Dirac semimetals are stable 3D electron systems with bulk Dirac nodes protected by crystalline symmetries~\cite{Young:2012, Steinberg:2014, YangNagaosa:2014}. In DSM, owing to the simultaneous presence of time reversal and space inversion symmetries, the bulk energy bands are Kramers degenerate locally at each $\mathbf{k}$ ($E_{n,\sigma}(\mathbf{k})=E_{n,-\sigma}(\mathbf{k})$). The Kramers degeneracy ensures that an accidental crossing between valence and conduction bands engenders a four-fold degenerate Dirac node. Such four-fold degenerate nodes in the bulk energy spectrum can be stable only in the presence of additional symmetries, such as uniaxial discrete crystal rotation symmetries $C_n$~\cite{YangNagaosa:2014}. This can be contrasted with three dimensional topological Weyl semimetals (WSM)~\cite{Murakami1:2007, Wan:2011, Xu:2011, Burkov:2011}, where (two-fold degenerate) Weyl nodes in the bulk energy spectrum are stable due to the existence of a non-zero Chern number invariant associated with each Weyl node. In DSM the simultaneous presence of time reversal and space inversion symmetry ensures that the Chern number vanishes for each Dirac node, which can in turn be thought as the superposition of a pair of Weyl nodes with equal and opposite chirality.

In recent studies several materials have been theoretically proposed to be topological DSMs~\cite{Wan:2011, Schoop:2015, Young:2012, Wang:2012, Wang:2013}. On the experimental side, Cd$_3$As$_2$~\cite{{Neupane:2014},{Borisenko:2014}, {Neupane:2015}, {Yi:2014}, {Liu1:2014}, {Jeon:2014}, {He:2014}, Liang:2015} and Na$_3$Bi~\cite{Liu2:2014, Xu:2015, Kushwaha:2015} have been experimentally confirmed to support three dimensional bulk Dirac nodes with linear energy spectrum. In this work we take a system with a pair of four-fold degenerate Dirac nodes on a high symmetry axis (which we choose as the $k_z$ axis as the axis of $C_n$ crystal rotation symmetry) as a prototypical topological DSM. The recently discovered DSMs, Cd$_3$As$_2$ and Na$_3$Bi, are both thought to be in this class. This class of DSMs are topological because, since the Dirac points appear on the high symmetry axis at $(k_x,k_y,k_z)=(0,0,\pm k_{z0})$ the system is invariant under a discrete rotation symmetry $C_4$ about the $k_z$ axis, which is responsible for the stability of the Dirac points. The $C_4$ discrete rotation symmetry about the $k_z$ axis allows one to define an additional integer topological invariant (mirror Chern number) on the $k_z=0$ plane \cite{YangNagaosa:2014}. The mirror symmetry appears here as a result of the combination of space inversion symmetry and $\pi$ rotation about the $k_z$ axis, which follows from the existence of the $C_4$ symmetry.

Although topological DSMs have certain non-trivial topological properties such as surface Dirac fermions and zero energy Fermi loops, topological thermoelectric response such as anomalous Hall and Nernst effects, which depend on non-zero momentum space integrals of Berry curvature across surfaces in the Brillouin zone, must vanish in the limit of zero magnetic field, because of the existence of time reversal symmetry. Since anomalous Hall and Nernst conductivities arise from the transverse current response (odd under time reversal) to an applied longitudinal electric field and temperature gradient (even under time reversal), independent of an applied magnetic field (odd under time reversal), it follows that the anomalous conductivities must vanish in systems that preserve time reversal symmetry. This can also be understood from the fact that the Chern number of the Dirac nodes in a DSM, which measures the flux of the Berry curvature over closed surfaces around the Dirac node, is identically zero, and thus, in the absence of a magnetic field, the net flux of the Berry curvature vanishes everywhere in the Brillouin zone. In the presence of a magnetic field, however, time reversal symmetry is broken, and topological DSMs reduce to WSMs, evincing anomalous Hall and Nernst response, superimposed over the conventional conductivities which must also be present because of a non-zero magnetic field.

In this paper we focus on the Nernst effect (conventional as well as anomalous) in topological Dirac semimetals for small magnetic fields (of the order of a few Tesla) and small chemical potential within the framework of Boltzmann theory in the relaxation time approximation. Using typical parameters~\cite{XiongEPL} for the scattering time $\tau\sim 0.1ps$ and $\omega_c\sim 1meV$, we find $\omega_c\tau\ll 1$, thus justifying our use of Boltzmann theory. We also make the standard assumption that any perturbation in the system decays exponentially with a relaxation time $\tau$, which is valid for small $(f_k-f_0)$ with $f_k$ ($f_0$) the perturbed (equilibrium) distribution function.
In our calculations we find that the conventional Nernst response is small due to Sondheimer cancellation, and can be of either sign depending on temperature, but the anomalous Nernst response is large and positive because of the peak in the Berry curvature in the limit of small chemical potentials. At low temperatures, the behavior of the total Nernst coefficient is characterized by an almost step like profile at $\mathbf{B}=0$.  However, exactly at $\mathbf{B}=0$ there is no Nernst signal from either conventional or anomalous contributions, because of the restoration of time reversal symmetry. The measured low-field Nernst coefficient, thus, is dominated by the anomalous Nernst effect, at least in the limit of small temperatures. Our results have direct experimental relevance for Nernst and thermoelectric measurements on the available topological Dirac semimetals Cd$_3$As$_2$ and Na$_3$Bi. Very recently, an anomalous Nernst signal has been reported in thermoelectric experiments on the Dirac semimetal Cd$_3$As$_2$ by the Princeton group~\cite{TianLiang:2016}, with a step-like profile at low magnetic fields very similar to our prediction.  For related work on thermoelectric response, although not for topological DSMs with a pair of Dirac points as appropriate for Cd$_3$As$_2$ and Na$_3$Bi, see Ref.~[\onlinecite{Sharma2:2016}] and Ref.~[\onlinecite{Lundgren:2014}].

Additionally in this paper we also discuss Nernst response of an inversion asymmetric WSM. Note that both the DSM and the inversion broken WSM are TR invariant systems, so a non-zero anomalous Nernst coefficient should be absent in the absence of a magnetic field. However in the presence of a magnetic field a DSM as well as an inversion broken WSM should show an anomalous (induced by Berry curvature) as well as a conventional Nernst response, both of which vanish in the limit of zero magnetic field. The Nernst effect in inversion broken WSM is thus expected to be similar to that in topological DSMs. However, we find that in the inversion broken WSM the conventional Nernst coefficient dominates over the anomalous contribution which is opposite to the case in DSMs. Our calculations on Nernst response in DSM and inversion broken WSM are directly relevant to recent experiments in CD$_3$As$_2$ and NbP respectively~\cite{TianLiang:2016,Sarah}.

This paper is organized as follows. In Sec. II we introduce the low energy Hamiltonian appropriate for a topological DSM with a pair of Dirac points on a high symmetry axis. This model should serve as an effective description for the available topological DSM Cd$_3$As$_2$. In Sec. III we briefly sketch the derivation of Nernst conductivity within Boltzmann theory description in relaxation time approximation in the presence of a non-trivial Berry curvature. In the presence of a non-zero magnetic field, the Nernst response of topological DSMs comprise conventional as well as anomalous components, which are then described in Sec. IV and Sec. V, respectively. Our central results, plots for the total Nernst conductivity (conventional as well as anomalous) as a function of the applied magnetic field at several different temperatures are displayed in Fig. 5. In Sec VI we additionally discuss the Nernst response of an inversion asymmetric WSM and point out the salient experimental features. In Sec VII we compare our findings to that of the recent experiments on Nernst response in Dirac semimetals and inversion broken Weyl semmetals. 
We end with a brief discussion and conclusion in Sec. VIII.

\section{Hamiltonian for topological DSM}
\begin{figure}
\centering
\includegraphics[scale=0.2]{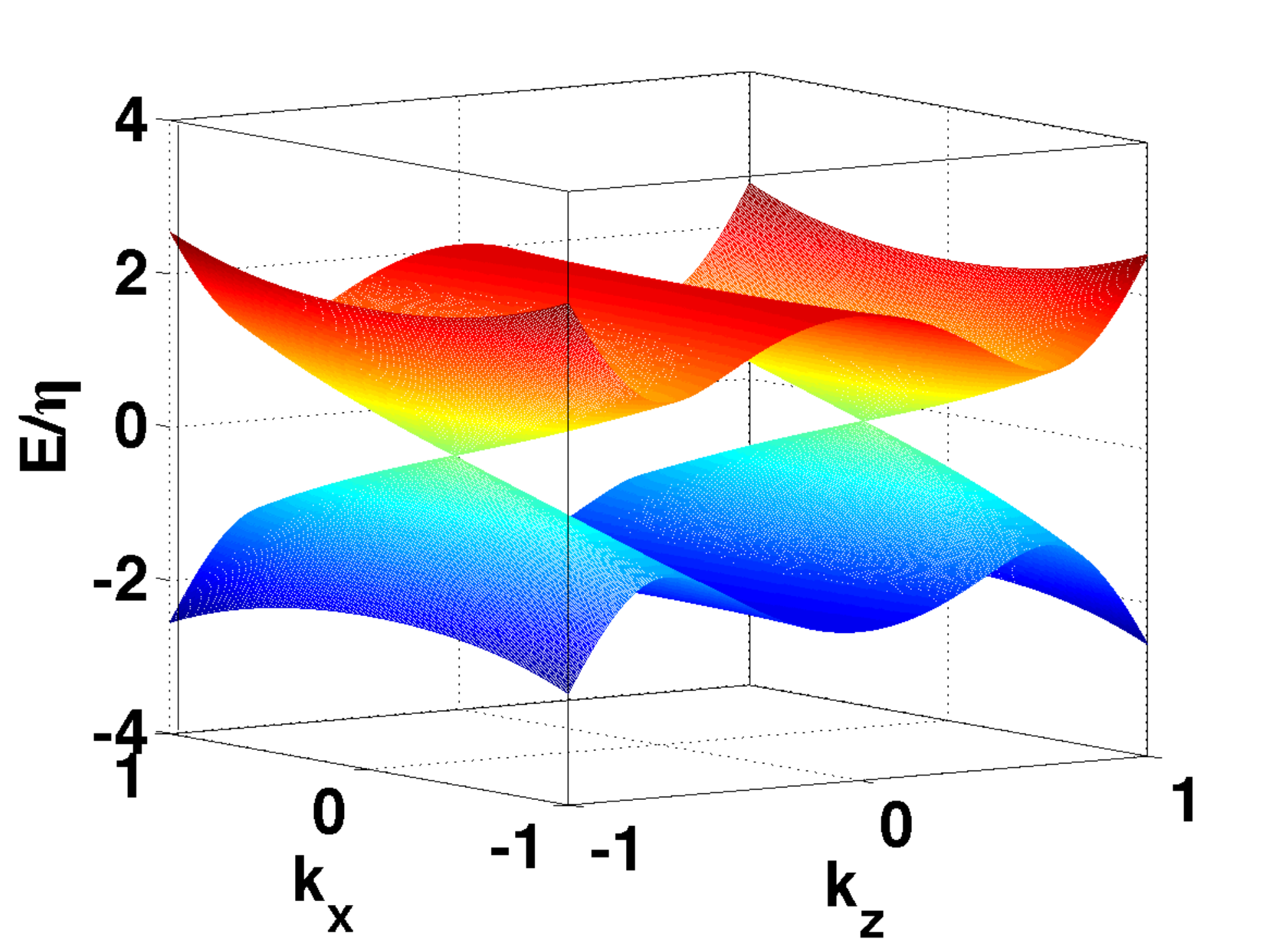}
\includegraphics[scale=0.2]{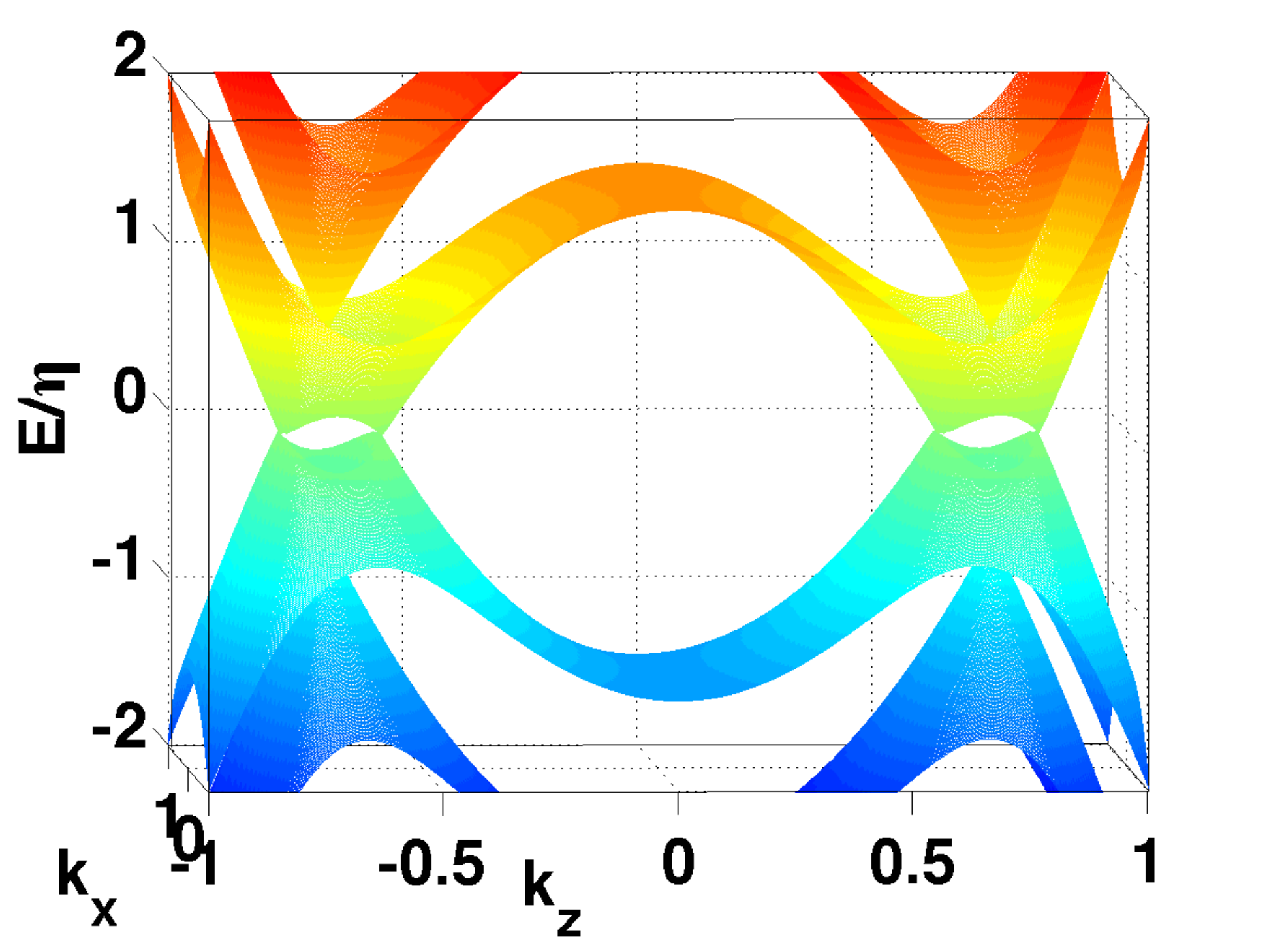}\\
\includegraphics[scale=0.3]{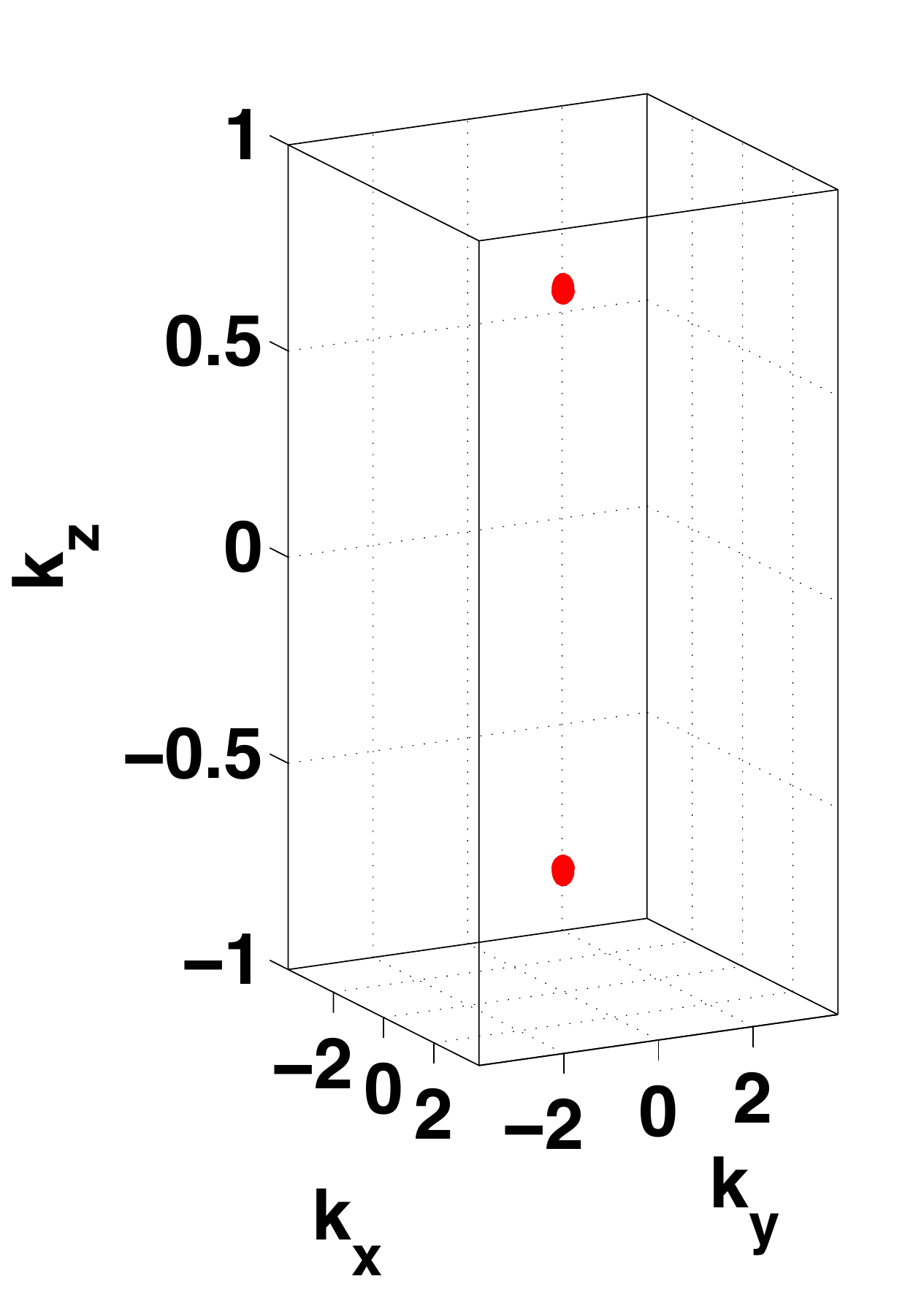}
\includegraphics[scale=0.3]{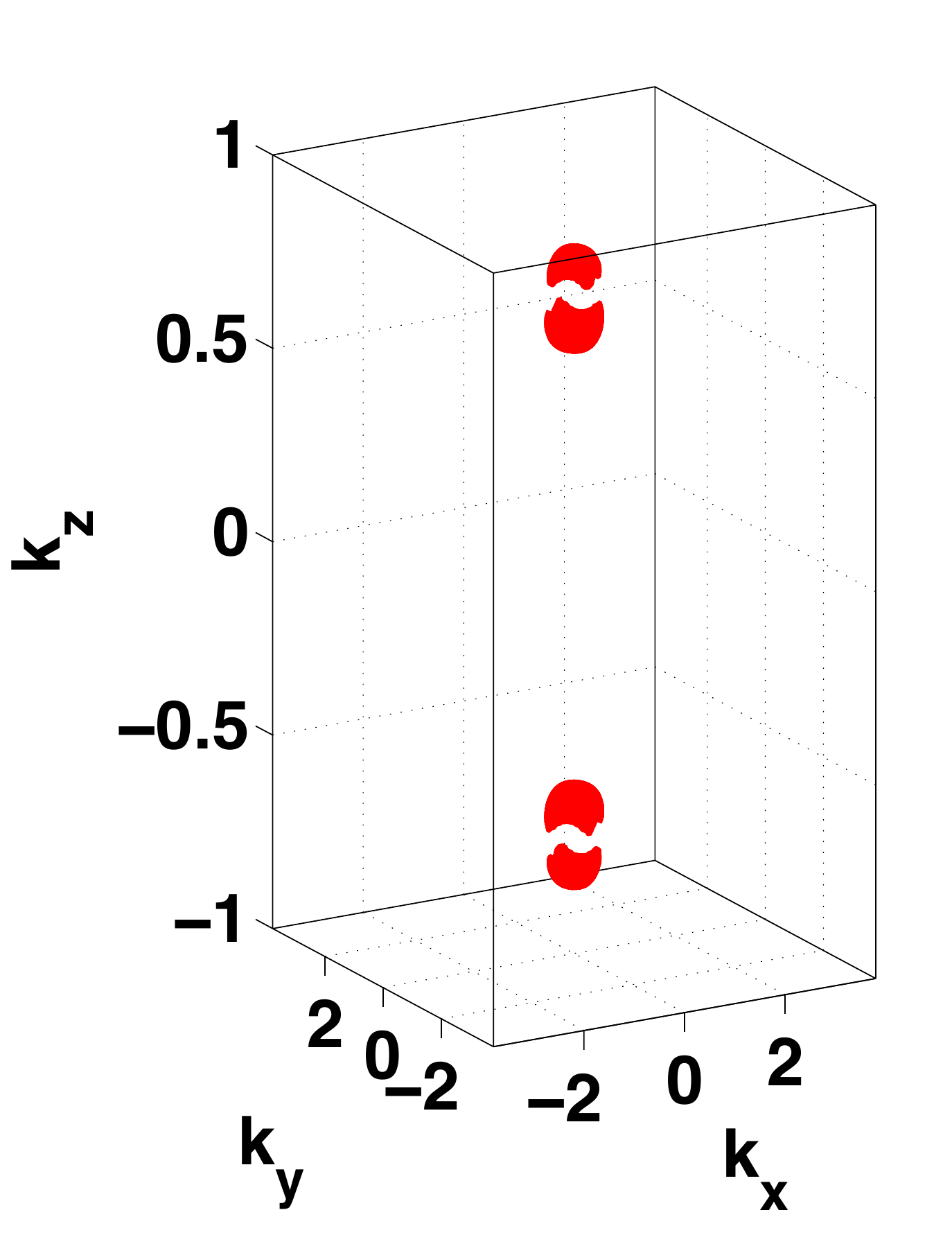}
\caption{\textit{Left panel:}  Band structure of the Dirac semimetal given by Eq.~\ref{Eqn_H_k_DSM} consisting of two four-fold degenerate Dirac points at $(0,0,\pm\sqrt{m_0/m_1})$. \textit{Right panel:} The spin-degeneracy is lifted by a magnetic field producing a total of four doubly-degenerate Weyl points located at $(0,0,\pm\sqrt{(\pm b_z + m_0)/m_1})$. The parameters~\cite{SatoPRL} used were $m_2=-\eta/5$, $m_0=-2\eta$, $m_1=-4\eta$, $\beta=-\eta/5$, $\gamma=\eta$, and $b_z=0$ ($b_z=3\eta/5$) for the (left) right panels. \textit{Bottom panels:} Fermi surfaces for the doped Dirac (Weyl) metals on the left (right), when $b_z=0$ ($b_z=3\eta/5$), for $\mu=0.16\eta$. The parameter $\eta$ was chosen to be $\eta=50meV$.}
\label{Fig_DSM_bands}
\end{figure}

\begin{figure}
\centering
\includegraphics[scale=0.33]{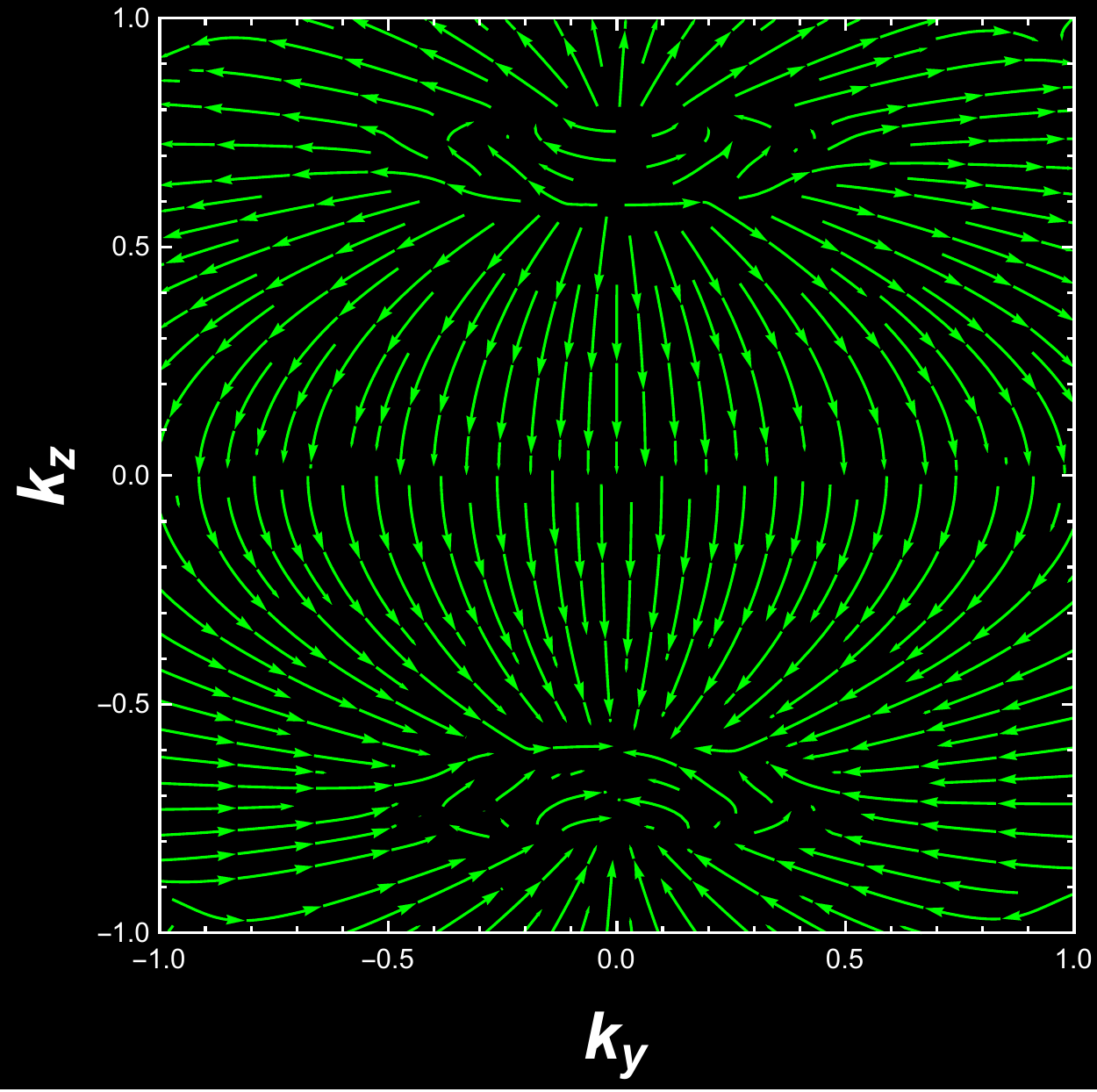}
\includegraphics[scale=0.33]{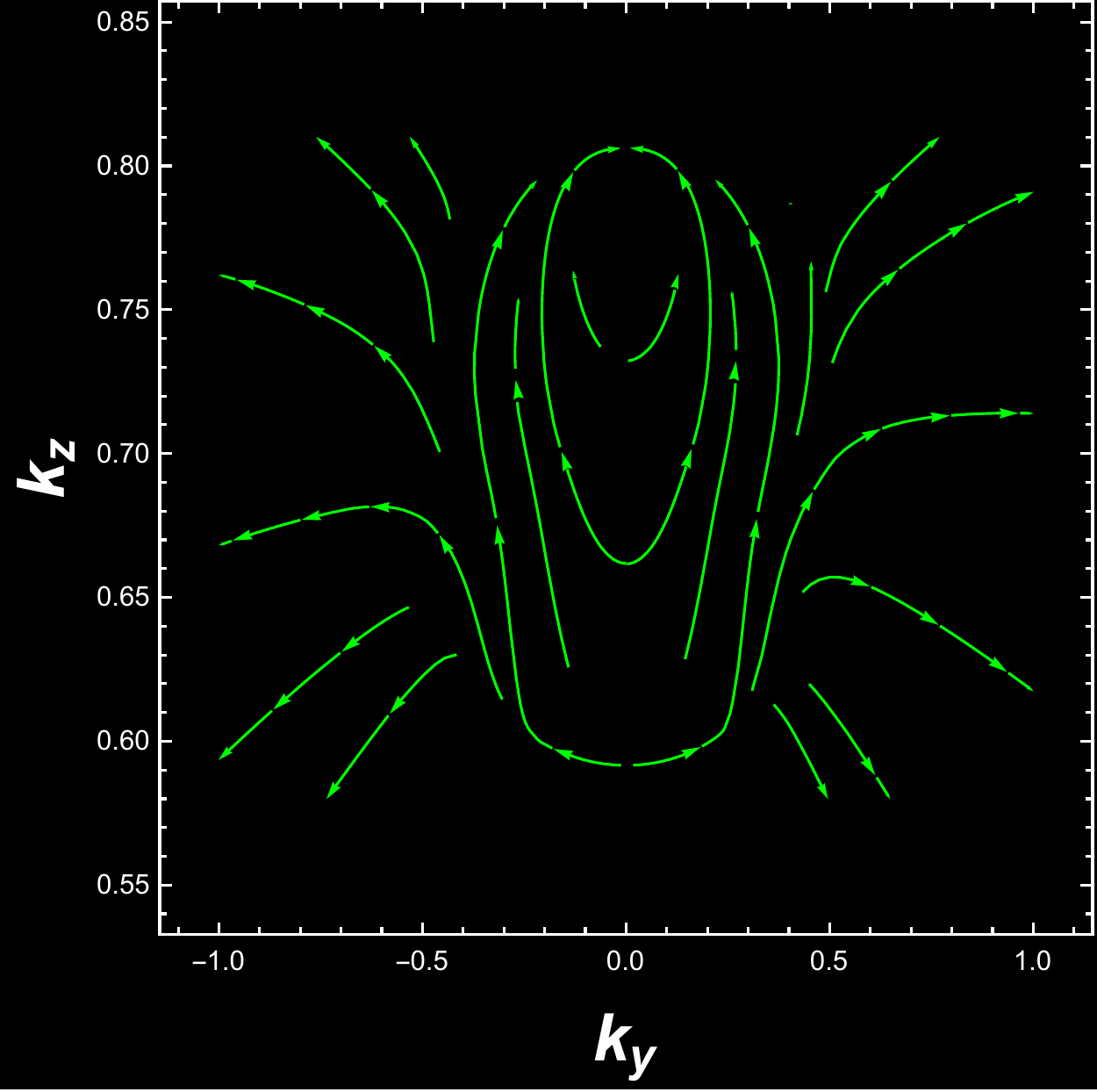}
\caption{Berry curvature flux for the topological DSM described by the Hamiltonian given in Eq.~\ref{Eqn_H_k_DSM}, in the presence of a magnetic field (in the presence of a magnetic field the system is effectively a Weyl semimetal). The plot on the left shows the Berry curvature in the plane $k_x=0$ suggesting a sink and source of Berry flux near each Dirac point at $(0,0,\pm\sqrt{m_0/m_1})$. On the right, we have the Berry curvature distribution zoomed in near a single Dirac point $(0,0,+\sqrt{m_0/m_1})\approx (0,0,0.7)$ (now split into two Weyl points at $\approx (0,0,0.6)$, and $\approx (0,0,0.8)$ when $b_z\neq 0$). A similar distribution (not shown specifically by zooming) exists around the other Dirac point $(0,0,-\sqrt{m_0/m_1})$.}
\label{Fig_Berry_curv}
\end{figure}

The effective low energy Hamiltonian for the Dirac semimetal Cd$_3$As$_2$, in the basis $|s, \uparrow\rangle$, $|p_x+ip_y, \uparrow\rangle$, $|s,\downarrow\rangle$, $|p_x-ip_y,\downarrow\rangle$ can be written as~\cite{YangNagaosa:2014,Hashimoto:2016}
\begin{eqnarray}
H_{\mathbf{k}} &= a(\mathbf{k})\sigma_zs_0 + b(\mathbf{k})\sigma_xs_z  + c(\mathbf{k})\sigma_ys_0 \nonumber\\
&+ d(\mathbf{k})\sigma_xs_x  + e(\mathbf{k})\sigma_xs_y
\label{Eqn_H_k_DSM}
\end{eqnarray}
In Eq.~\ref{Eqn_H_k_DSM}, $\sigma$ and $s$ are Pauli matrices representing the orbital degree of freedom and spin degree of freedom respectively. The matrix $s_0\equiv I_2$ is the two-dimensional identity matrix in spin space. The functions $a(\mathbf{k})-e(\mathbf{k})$ are defined as
\begin{align}
a(\mathbf{k})&=m_0-m_1 k_z^2-m_{2}(k_x^2+k_y^2),\label{Eq_a_DSM}\\
b(\mathbf{k})&=\eta k_x,\label{Eq_b_DSM}\\
c(\mathbf{k})&=-\eta k_y,\label{Eq_c_DSM}\\
d(\mathbf{k})&=(\beta + \gamma) k_z (k_y^2 - k_x^2),\label{Eq_d_DSM}\\
e(\mathbf{k})&=-2(\beta - \gamma) k_z k_x k_y,\label{Eq_e_DSM}
\end{align}
The parameters $m_0$, $m_1$, $m_2$, $\eta$, $\beta$ and $\gamma$ depend on the material. For example for Cd$_3$As$_2$ ab-inito calculations upto order $k^2$ yield $m_0=.02eV$, $m_1=-18.77eV \AA^2$, $m_2=-13.5 eV \AA^2$, $\eta=0.89 eV\AA$~\cite{Cano:2016}.
This Hamiltonian produces two Dirac points at $\mathbf{K}=(0,0,\pm\sqrt{m_0/m_1})$ where the energy dispersion exactly vanishes. Fig.~\ref{Fig_DSM_bands} shows the band structure for the prototype DSM obtained by numerically diagonalizing Eq.~\ref{Eqn_H_k_DSM}. The effect of an external magnetic field $\mathbf{B}$, coupling to the spin degree of freedom can be now introduced by adding the Zeeman term $H_Z = b_z \sigma_0s_z$ in the Hamiltonian, where $b_z=-\boldsymbol{\mu}\cdot{\mathbf{B}}$, $\boldsymbol{\mu}$ being the spin-magnetic moment, $\boldsymbol{\mu}=-\mu_B g_s \mathbf{s}/\hbar$.
{For typical experimental parameters for NbP and Na$_3$Bi ($g_s\sim 20-40$ and $m*\sim 0.11m_0$)~\cite{ShekharNature, XiongEPL} we find that the Zeeman energy scale is $\sim 2meV$ and the typical orbital energy scale $\hbar \omega_c\sim 0.9 meV$. Therefore for the sake of simplicity in the following we ignore the orbital coupling of the magnetic field to the Dirac electrons.}
With the applied magnetic field the Hamiltonian now produces a TR broken Weyl semimetal, with four Weyl points located at $(0,0,\pm\sqrt{(\pm b_z + m_0)/m_1})$. Each Weyl node now carries a non-trivial Chern number, which is also its chirality quantum number. Fig.~\ref{Fig_DSM_bands} also shows the band structure for the TR broken Weyl semimetal. Near half-filling, the Fermi surface for a Dirac semimetal consists of two disconnected spheres, as shown in Fig.~\ref{Fig_DSM_bands}, for $\mu>0$. The Zeeman field then splits each sphere into two disconnected surfaces around each Weyl point. Although the topological DSM described by the Hamiltonian in Eq.~\ref{Eqn_H_k_DSM} is characterized by a vanishing Berry curvature, the topological WSM described by $H=H_{\mathbf{k}}+H_Z$ has a nontrivial profile of Berry curvature in the Brillouin zone.
The expression for the Berry curvature is given by~\cite{Xiao:2010},
\begin{eqnarray}
\Omega_{ab}^n = i\sum\limits_{n\neq m}\frac{\langle n|\partial H/\partial k_a|m\rangle \langle m|\partial H/\partial k_b|n\rangle -(a\leftrightarrow b )}{(\epsilon_n - \epsilon_m)^2} \nonumber \\
\label{Eq_Berry_curv}
\end{eqnarray}
The Hamiltonian $H=H_{\mathbf{k}}+H_Z$ produces four bands which we have labeled by the index $n$ in the above expression. Also, $|n\rangle$ is a Bloch eigenstate of the Hamiltonian $H$ with eigenvalue $\epsilon_n$. In Fig.~\ref{Fig_Berry_curv} we have plotted the Berry curvature distribution for the DSM with Zeeman coupling described by the Hamiltonian $H=H_{\mathbf{k}}+H_Z$.

\section{Nernst effect in the presence of Berry curvature}
The Nernst effect refers to the generation of a transverse electric field in the presence of a longitudinal temperature gradient. Conventionally, the Nernst effect can occur only in the presence of an external magnetic field, which provides a transverse velocity to the electrons by the Lorentz force. However, a non-trivial Berry curvature $\mathbf{\Omega}$, can also give rise to a Nernst response as a result of an anomalous velocity term~\cite{Niu:2006}. In the presence of an external electric field $\mathbf{E}$ and a temperature gradient $-\nabla T$, one can write the following linear response relations for the charge current $\mathbf{J}$ and thermal current $\mathbf{Q}$:
\begin{eqnarray}
\left( \begin{array}{c}
\mathbf{J} \\
\mathbf{Q}  \\
\end{array} \right) =\left( \begin{array}{cc}
\hat{\sigma} & \hat{\alpha} \\
\hat{\bar{\alpha}} & \hat{\kappa}  \\
\end{array} \right) \left( \begin{array}{c}
\mathbf{E} \\
-\mathbf{\nabla} T  \\
\end{array} \right)
\label{Eq_thermal}
\end{eqnarray}
The tensors $\hat{\bar{\alpha}}$ and $\hat{\alpha}$ are related to each other by Onsager's relation: $\hat{\bar{\alpha}}=T\hat{\alpha}$. In the absence of charge current ($\mathbf{J}=0$), we have $\mathbf{E}=\hat{\sigma}^{-1}\hat{\alpha}\nabla T$. The Nernst coefficient $\nu$ can be derived to be
\begin{align}
\nu =   \frac{E_y}{(-dT/dx)} = \frac{\alpha_{xy}\sigma_{xx} - \alpha_{xx}\sigma_{xy}}{\sigma_{xx}^2 + \sigma_{xy}^2},
\label{Eq_nernst}
\end{align}
Berry curvature significantly contributes to the conductivities $\boldsymbol{\sigma}$ and $\boldsymbol{\alpha}$. In the presence of Berry curvature $\mathbf{\Omega}_{\mathbf{k}}$, the semi-classical equation of motion for an electron takes the form\cite{Niu:2006, Sundurum:1999} $\mathbf{\dot{r}} = \frac{1}{\hbar} \frac{\partial\epsilon(\mathbf{k})}{\partial\mathbf{k}} + \frac{\mathbf{\dot{p}}}{\hbar}\times\mathbf{\Omega}_{\mathbf{k}}$. The first term is the familiar relation between semi-classical velocity $\mathbf{\dot{r}}$ and the band energy dispersion $\epsilon(\mathbf{k})$. The second term is the anomalous transverse velocity term originating from $\mathbf{\Omega(\mathbf{k})}$.  In the presence of electric and magnetic fields we also have the standard relation $\mathbf{\dot{p}} = e\mathbf{E} + e\mathbf{\dot{r}}\times\mathbf{B}$. These two coupled equations for $\mathbf{\dot{r}}$ and $\mathbf{\dot{p}}$ can be solved together to obtain~\cite{Duval:2006, Son:2012}
\begin{align}
&\mathbf{\dot{r}} = D(\mathbf{B},\Omega_{\mathbf{k}}) \left(\mathbf{v}_{\mathbf{k}} + \frac{e}{\hbar} (\mathbf{E}\times\Omega_{\mathbf{k}}) + \frac{e}{\hbar}(\mathbf{v}_{\mathbf{k}}\cdot\Omega_{\mathbf{k}})\mathbf{B}\right)\\
&\mathbf{\dot{p}} = D(\mathbf{B},\Omega_{\mathbf{k}}) \left(e\mathbf{E} + \frac{e}{\hbar} (\mathbf{v}_{\mathbf{k}}\times B) + \frac{e^2}{\hbar}(\mathbf{E}\cdot\mathbf{B})\mathbf{\Omega_{\mathbf{k}}}\right)
\label{pdot_eqn}
\end{align}
where $D(\mathbf{B},\Omega_{\mathbf{k}}) = (1+e (\mathbf{B}\cdot\Omega_{\mathbf{k}})/\hbar)^{-1}$.

Using the semi-classical Boltzmann equations in the presence of a non-zero electric and magnetic field and a Berry curvature, one can derive the following thermoelectric  and charge conductivity tensors ($\boldsymbol{\sigma}$ and $\boldsymbol{\alpha}$) which include contributions from the $\mathbf{B}$ and $\boldsymbol{\Omega}_{\mathbf{k}}$~\cite{Sharma2:2016, Lundgren:2014, Kim:2014, Son:2013}.
\begin{align}
\sigma_{xx} &= -e^2\int{[d\mathbf{k}] v_x^2\tau \left(-\frac{\partial f_{eq}}{\partial\epsilon}\right) \left(c_x-D \right)}
\label{sxx2}\\
\sigma_{xy} &= {-e^2}\int{[d\mathbf{k}] (v_y^2 c_y\ + v_xv_y(c_x-D)) \tau\left(-\frac{\partial f_{eq}}{\partial\epsilon}\right)} \nonumber \\
& + \frac{e^2}{\hbar} \int{[d\mathbf{k}] \mathbf{\Omega}_z f_{0}}
\label{syx2}
\end{align}
\begin{align}
\alpha_{xx} &= e\int{[d\mathbf{k}] v_x^2\left(\tau\frac{\epsilon-\mu}{T} \left(-\frac{\partial f_{eq}}{\partial\epsilon}\right) \left(c_x-D \right) \right)}
\label{axx2}\\
\alpha_{xy} &= e\int{[d\mathbf{k}] (v_y^2 c_y + (c_x-D)v_xv_y) \left(\tau\frac{\epsilon-\mu}{T} \left(-\frac{\partial f_{eq}}{\partial\epsilon}\right) \right)}\nonumber \\
&  + {\frac{k_Be}{\hbar} \int{[d\mathbf{k}] \mathbf{\Omega}_z s_{\mathbf{k}}}}
\label{ayx2}
\end{align}
where $v_x\equiv\hbar^{-1}\partial \epsilon_{\mathbf{k}}/\partial k_x$, and $v_y\equiv\hbar^{-1}\partial \epsilon_{\mathbf{k}}/\partial k_y$ are the band velocities, $\epsilon_F$ is the Fermi energy, $\tau$ is the scattering time, $[d\mathbf{k}]\equiv \frac{d^3\mathbf{k}}{(2\pi)^3}$, $f_0$ is the Fermi-Dirac distribution,  $s_{\mathbf{k}} = -f_{0}\log f_{0} - ((1-f_{0}) \log(1-f_{0}))$ is entropy density for the free electron gas. The scattering time $\tau$ usually has a non-trivial energy and momentum dependence. In this work we have assumed $\tau$ to be a phenomenological constant, which suffices for our discussion and does not change our qualitative results. The correction factors $c_x$, $c_y$, and $D$ in Eq.~\ref{sxx2}-\ref{ayx2} have lengthy expressions and have been discussed elsewhere~\cite{Sharma2:2016}. In the absence of Berry curvature $D\rightarrow 1$, and $c_x-D\rightarrow -1$ (up to zeroth order in $\mathbf{B}$). It is important to note that the longitudinal conductivities ($\sigma_{xx}$, $\alpha_{xx}$) are also modified from their standard Boltzmann expressions due to Berry curvature corrections. If these corrections can be ignored, then Eq.~\ref{sxx2},~\ref{axx2} reduce to the following~\cite{Ziman}
\begin{align}
&\sigma_{xx} = e^2\int{[d\mathbf{k}] v_x^2 \tau \left(-\frac{\partial f_{eq}}{\partial\epsilon}\right)},\label{Eq_sxx_conv}\\
&\alpha_{xx} = -\frac{e}{T}\int{[d\mathbf{k}] v_x^2 \tau (\epsilon-\mu) \left(-\frac{\partial f_{eq}}{\partial\epsilon}\right)}, \label{Eq_axx_conv}
\end{align}
Similarly, if the Berry curvature corrections to the conventional $\mathbf{B}$-dependent conductivities are ignored, then Eq.~\ref{syx2},~\ref{ayx2} reduce to~\cite{Niu:2006, Ziman}
\begin{align}
&\sigma_{xy} = -\frac{e^3\tau^2B}{\hbar} \int{[d\mathbf{k}] \(-\frac{\partial f_0}{\partial\epsilon}\)\(\frac{v_x^2\partial^2\epsilon}{\partial k_y^2}-\frac{v_xv_y\partial^2\epsilon}{\partial k_x\partial k_y}\)} \nonumber \\ & + \frac{e^2}{\hbar} \int{[d\mathbf{k}] \mathbf{\Omega}_z f_{0}}
\label{Eq_sxy_conv}
\end{align}
\begin{align}
&\alpha_{xy} = \frac{e^3\tau^2B}{T\hbar} \int{[d\mathbf{k}] (\epsilon-\mu)\(-\frac{\partial f_0}{\partial\epsilon}\)\(\frac{v_x^2\partial^2\epsilon}{\partial k_y^2}-\frac{v_xv_y\partial^2\epsilon}{\partial k_x\partial k_y}\)} \nonumber \\
&  + {\frac{k_Be}{\hbar} \int{[d\mathbf{k}] \mathbf{\Omega}_z s_{\mathbf{k}}}}
\label{Eq_axy_conv}
\end{align}
\section{Conventional Nernst response}
The conventional Nernst coefficient can be deduced by using Eqs.~\ref{Eq_sxx_conv}-\ref{Eq_axy_conv} in the limit $\boldsymbol{\Omega}_{\mathbf{k}}\rightarrow 0$, and the definition of $\nu$ (Eq.~\ref{Eq_nernst}). In conventional metals, the quasiparticle Nernst coefficient is usually small as a result of Sondheimer cancellation~\cite{Sondheimer,Wang:2001}. For example, the Nernst coefficient $\nu/B$ is $3.9 nV/KT$ for Al, and $-21.6 nV/KT$ for Cu~\cite{Rumbo:76, Behnia:04}. In the limit of small $\mu$, the conventional Nernst coefficient for a linearized Dirac Hamiltonian ($\epsilon_{\mathbf{k}}=\hbar v_F \boldsymbol{\sigma}\cdot\mathbf{k}$) can be derived to be~\cite{Sharma2:2016}
\begin{equation}
\vartheta_0 = -\frac{\pi^2}{3}\frac{k_B^2 T}{e}\frac{eBv_F^2\tau}{\hbar},
\end{equation}
where $\tau$ parametrizes the scattering time. If $\tau$ is large, then the Nernst coefficient can also be parametrically large even in the presence of Sondheimer's cancellation.  For our model and the chosen parameters~\cite{Cano:2016} for Cd$_3$As$_2$, using Eqs.~\ref{Eq_sxx_conv}-\ref{Eq_axy_conv} with $\boldsymbol{\Omega}_{\mathbf{k}}\rightarrow 0$ and $\tau\sim0.1 ps$~\cite{XiongEPL},  we found that the conventional Nernst coefficient is at least one order of magnitude smaller than the anomalous Berry curvature dependent response, due to the peaks in the Berry curvature near $\mu=0$ at the four nodal Weyl points (see Fig. 3).
Importantly, the sign of the conventional Nernst coefficient is not directly related to the sign of the dominant charge carriers in the material. It can be either positive or negative for an electron or a hole-like Fermi surface, depending on the detailed Fermi surface topology~\cite{Wang:2001, Zhang:2010}. Our model produces a negative conventional Nernst signal (for positive $b_z$) at low temperatures ($T\sim 10K$), and a positive Nernst signal (for positive $b_z$) at higher temperatures ($T\sim 100K$). Further, electron ($\mu>0$) or hole ($\mu<0$) doping does not change the sign of the conventional Nernst coefficient.
\begin{figure}
\centering
\includegraphics[scale=0.35]{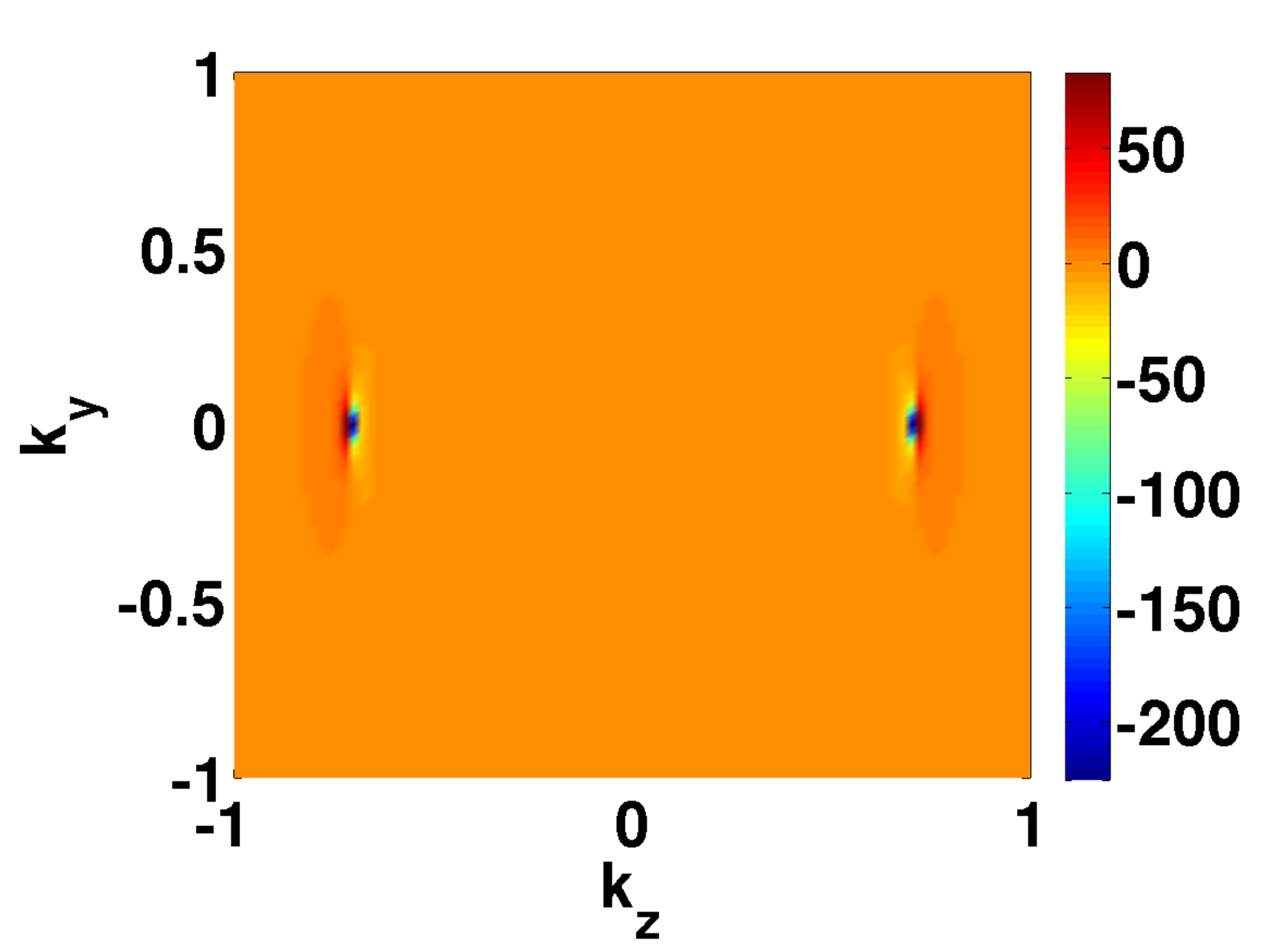}
\caption{Berry curvature $\Omega_z$ in the $k_x=0$ plane for the Weyl semimetal phase of the Hamiltonian in Eq.~\ref{Eqn_H_k_DSM}, for $b_z=\eta/5$. The Berry curvature peaks around the nodal (Weyl) points on the $k_z$ axis. For a small chemical potential, when the Fermi surface just encloses the Weyl points, the Nernst response is primarily dominated by the anomalous Berry curvature dependent contributions.  }
\label{Fig_Berry_plot}
\end{figure}

Though the conventional quasiparticle Nernst signal is known to be small, the Nernst effect has been used as a probe for high-$T_c$ cuprate superconductors, where vortex movement is well-known to give rise to a large \textit{positive} Nernst signal~\cite{Wang:2001, Behnia:2009}. This also forms the commonly used convention to assign a definite sign to a Nernst signal. We have followed this sign convention in our work.
\begin{figure}
\centering
\includegraphics[scale=0.35]{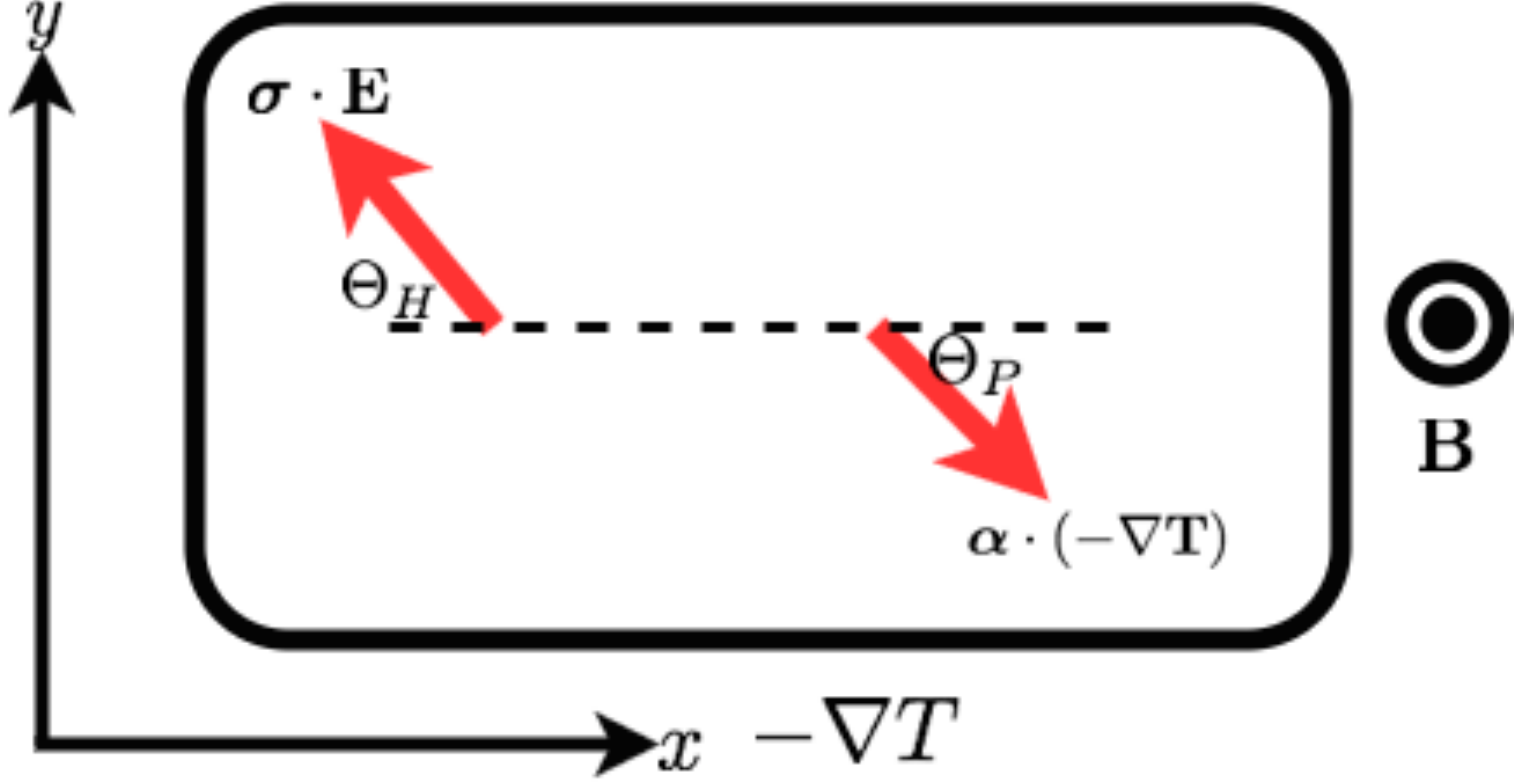}\\
\includegraphics[scale=0.35]{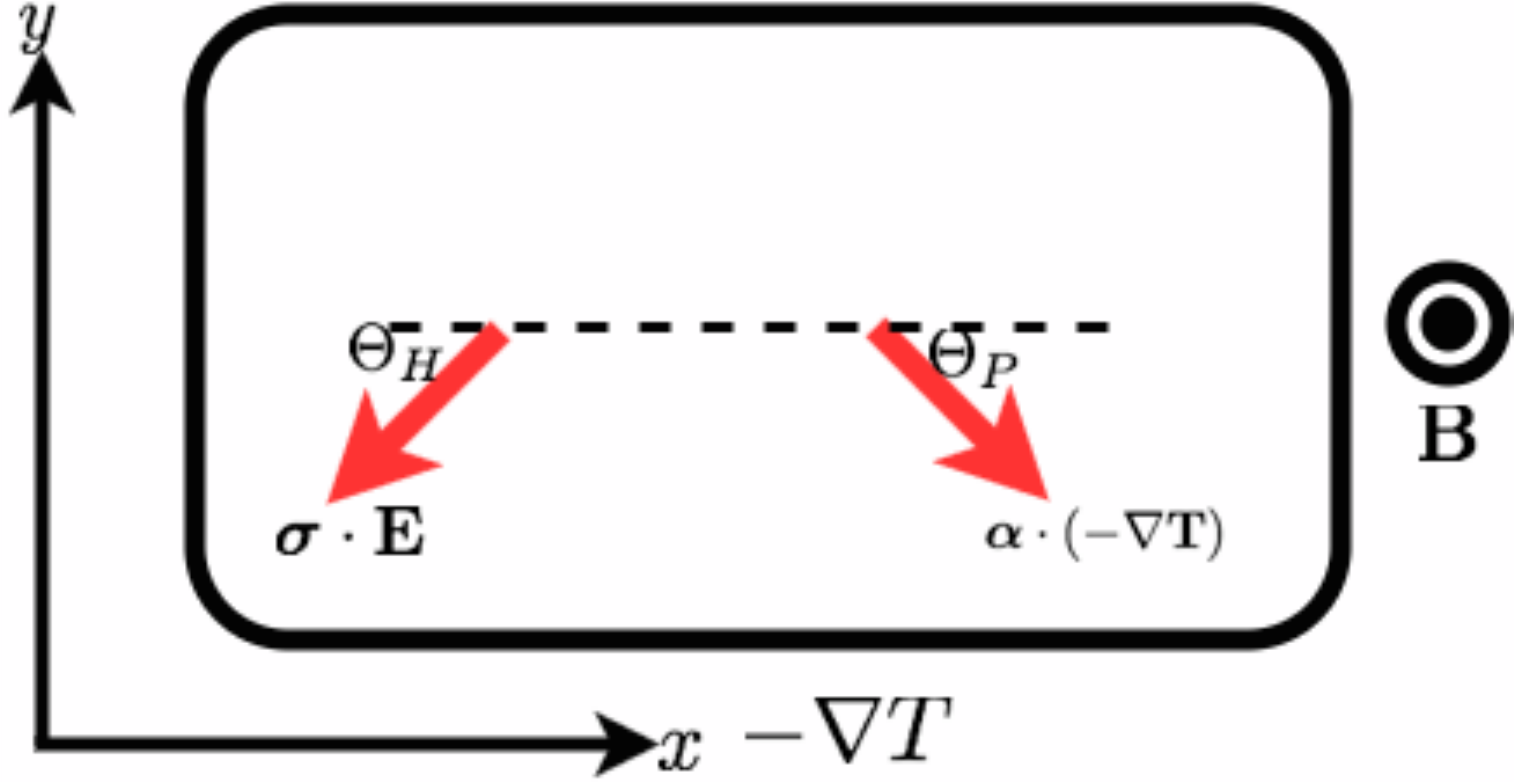}
\caption{\textit{Top panel:} Sondheimer's cancellation for conventional quasiparticle Nernst effect. The magnitudes of the Hall angle $\Theta_H$ and the Peltier angle $\Theta_P$ are close to each other, with $\text{sgn}(\Theta_H)=\text{sgn}(\Theta_P)$, resulting in a small Nernst signal in the presence of a longitudinal temperature gradient $-\nabla T$ and a perpendicular magnetic field $\mathbf{B}$. The red arrows represent the current direction due to the electric field and temperature gradient. \textit{Bottom panel:} The Hall and the Peltier angles no longer have the same signs for the anomalous Nernst response in a Dirac semimetal, resulting in no net Sondheimer's cancellation. The magnetic field breaks TR symmetry giving rise to a Weyl system with a measurable Nernst signal (electric field generated in the $y$ direction for a temperature gradient $-\nabla T$ in the $x$ direction), which is primarily anomalous response due to the peaking of the Berry curvature for small chemical potentials.}
\label{Fig_Nernst_setup}
\end{figure}

At low temperatures, the Mott relation gives $\alpha_{ij}$ as a derivative of $\sigma_{ij}$ with respect to the chemical potential~\cite{Ziman}. Specifically,
\begin{eqnarray}
\alpha_{ij} = -\frac{\pi^2}{3}\frac{k_B^2 T}{e}\frac{\partial\sigma_{ij}}{\partial\mu}
\label{Eq_Mott_relation}
\end{eqnarray}
The Mott relation (at least at low temperatures) remains valid for both conventional and anomalous conductivities.   Using the Mott relation, the Nernst coefficient $\nu$ can be derived to be
\begin{eqnarray}
\nu = -\frac{\pi^2}{3}\frac{k_B^2 T}{e} \frac{\partial \Theta_H}{\partial \mu},
\label{Eq_Nernst_Mott_relation}
\end{eqnarray}
where $\Theta_H=\sigma_{xy}/\sigma_{xx}$ is the Hall angle, in the limit $\sigma_{xy}\ll\sigma_{xx}$.  Expanding $\partial \Theta_H/\partial \mu$, we have
\begin{eqnarray}
\nu &= -\frac{\pi^2}{3}\frac{k_B^2 T}{e\sigma_{xx}^2} \left(\sigma_{xx}\frac{\partial\sigma_{xy}}{\partial\mu}-\sigma_{xy}\frac{\partial\sigma_{xx}}{\partial\mu} \right)\nonumber\\
&=\frac{\alpha_{xx}}{\sigma_{xx}}\left( \Theta_P - \Theta_H\right) ,
\label{Eq_Nernst_expanded}
 \end{eqnarray}
where $\Theta_P$ is the Peltier angle, with $\Theta_P = \alpha_{xy}/\alpha_{xx}$.
For a Dirac node having a spherical Fermi surface the longitudinal conductivity ($\sigma_{xx}$), which depends on the area of the Fermi surface, increases (decreases) for an electron (hole) doped system, with increasing $\mu$. The Hall conductivity ($\sigma_{xy}$), which correlates with the Fermi surface curvature is negative (positive) for electron (hole) doping.  However, $\sigma_{xx}>0$, and $\partial\sigma_{xy}/\partial\mu<0$ for both electron and hole-like Fermi surfaces. Hence the Hall and Peltier angles carry the same sign in Eq.~\ref{Eq_Nernst_expanded}. Sondheimer's cancellation~\cite{Sondheimer,Wang:2001,Behnia:2009} occurs when the angles $\Theta_H$ and $\Theta_P$ are close to each other in magnitude and have the same sign, $\text{sgn}(\Theta_H)=\text{sgn}(\Theta_P)$. This is the case in our calculation of the conventional Nernst response, resulting in a conventional Nernst coefficient much smaller in magnitude than the anomalous Nernst coefficient, which does not undergo Sondheimer cancellation.  This has also been illustrated in Fig.~\ref{Fig_Nernst_setup}, where the currents due to charge conductivity tensor $\boldsymbol{\sigma}$ and the Peltier coefficient $\boldsymbol{\alpha}$, oppose each other, in the case of conventional Nernst response.
\begin{figure}
\centering
\includegraphics[scale=0.5]{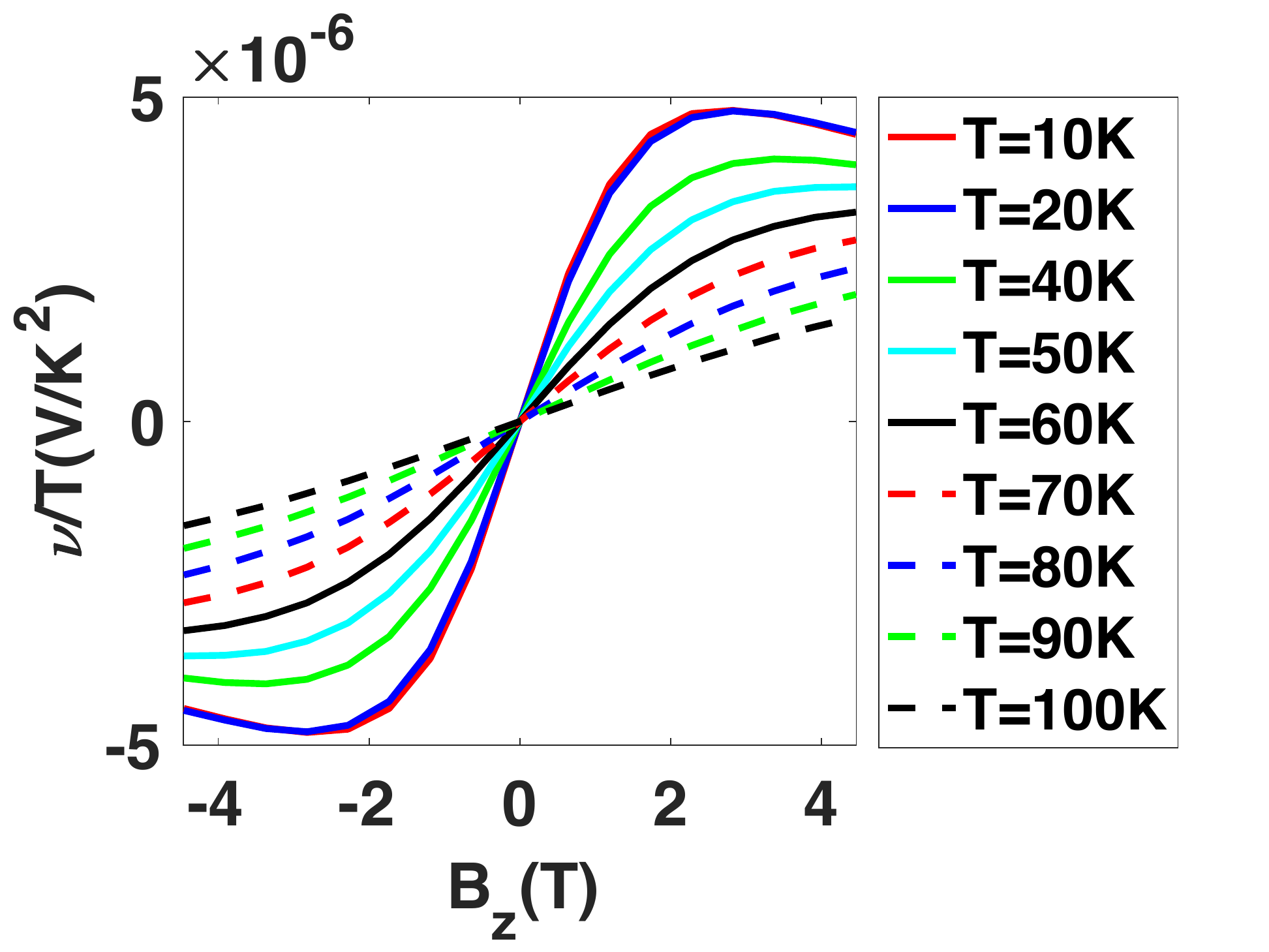}
\caption{Nernst coefficient ($\nu/T$) as a function of applied magnetic field $\mathbf{B}$, for the Dirac semimetal at $\mu=0.01eV$ (with an electron-like Fermi surface). The full Nernst coefficient has been plotted, although we find that the response is primarily dominated by the anomalous contribution which is an order of magnitude larger than the conventional contribution. At lower temperatures, the behavior is characterized by an almost step like profile near $\mathbf{B}=0$.  However, exactly at $\mathbf{B}=0$ there is no Nernst signal from either conventional or anomalous contributions. The estimated Nernst coefficient is of the order $\sim \mu V/K^2$. Recently, an anomalous Nernst signal has been reported in thermoelectric experiments on topological Dirac semimetal Cd$_3$As$_2$ by the Princeton group~\cite{TianLiang:2016}, with a step-like profile similar to above. The values of parameters were chosen from the ab-initio calculations~\cite{Cano:2016} for Cd$_3$As$_2$ DSM as given below Eq.~(6) and $\tau\sim 0.1ps$~\cite{XiongEPL}. }
\label{Fig_Nernst_full}
\end{figure}

\section{Anomalous Nernst response}
In the presence of Berry curvature ($\mathbf{\Omega}_{\mathbf{k}}$), Eq.~\ref{Eq_sxx_conv}-\ref{Eq_axy_conv} can be used to compute the Nernst coefficient. The anomalous transverse conductivities can be extracted to be
\begin{align}
&\sigma_{xy}^A = \frac{e^2}{\hbar} \int{[d\mathbf{k}] \mathbf{\Omega}_z f_{0}} ,\label{Eq_syx_anom}\\
&\alpha_{xy}^A = {\frac{k_Be}{\hbar} \int{[d\mathbf{k}] \mathbf{\Omega}_z s_{\mathbf{k}}}},
\label{Eq_ayx_anom}
\end{align}
 The quantity $\sigma_{xy}^A$ depends on the Berry curvature of the filled bands, but $\alpha_{xy}^A$ is a Fermi surface quantity, because $s_{\mathbf{k}}$ is zero for completely filled and empty bands. It is for this reason, that an insulator can give rise to an anomalous Hall response ($\sigma_{xy}^A$), but not $\alpha_{xy}^A$. The Dirac semimetal itself does not result in an anomalous Nernst signal, as the net flux of Berry curvature exactly vanishes everywhere in the Brillouin zone. Under the application of an external magnetic field $\mathbf{B}$, each Dirac node splits into two Weyl nodes, and near half-filling a Weyl semimetal is realized. The transition from a Dirac semimetal to a Weyl semimetal under the application of a magnetic field has been highlighted in Fig.~\ref{Fig_DSM_bands}. A Weyl semimetal has a non-trivial distribution of magnetic flux (as illustrated in Fig.~\ref{Fig_Berry_curv}) and Fig.~\ref{Fig_Berry_plot}, and an anomalous Nernst signal can thus be expected.

Unlike the conventional Hall conductivity ($\sigma_{xy}$), the anomalous Hall conductivity ($\sigma_{xy}^A$) has a different behavior with respect to changes in the chemical potential. The magnitude of the anomalous Hall conductivity peaks near the band-touching points at $\mu=0$, as in the vicinity of these points the Berry curvature is sharply peaked (Fig.~\ref{Fig_Berry_plot}). For small electron or hole doping, when $\mu\neq 0$, the Berry curvature effects reduce and the magnitude of $\sigma_{xy}^A$ decreases. The Mott relation (Eq.~\ref{Eq_Mott_relation}), thus produces opposite signs of $\alpha_{xy}^A$ for electron (positive sign) and hole doping (negative sign). 
The anomalous Peltier coefficient $\alpha_{xy}^A=0$ for an arbitrary $\mu$, if the underlying  quasiparticle dispersion is that of of an unbounded linearized spectrum of Weyl fermions, because then $\sigma_{xy}^A$ is robust to changes in the Fermi energy~\cite{Lundgren:2014, Sharma2:2016}.  Specifically, $\sigma_{xy}^A = \frac{e^2}{2\pi^2\hbar} k_0 $ for a simple linearized model of a Weyl semimetal with node separation given by $k_0$ in momentum space~\cite{Goswami1:2012, Burkov:2011, Burkov:2014}. However, for a physical Weyl semimetal with an ultraviolet cutoff, $\alpha_{xy}^A$ remains generically finite~\cite{Sharma2:2016}.

Now we note that for the anomalous conductivities ($\sigma_{xy}^A$, $\alpha_{xy}^A$), the Hall ($\Theta_H$) and Peltier ($\Theta_P$) angles, irrespective of their own  magnitudes (which may or may not be of the same order), have opposite signs of each other, $\text{sgn}(\Theta_H)=-\text{sgn}(\Theta_P)$ (in contrast to the case of conventional Nernst response where $\text{sgn}(\Theta_H)=\text{sgn}(\Theta_P)$). This suggests that the Sondheimer's cancellation does not take place, generating a measurable anomalous Nernst signal, stronger than the conventional quasiparticle Nernst signal. This feature has also been illustrated in Fig.~\ref{Fig_Nernst_setup}. The angles $\Theta_H$ and $\Theta_P$ carrying opposite signs result in a net non-zero electric field $E_y$ in the transverse direction.
The overall sign of the anomalous Nernst signal in the present case correlates with the sign of the anomalous Hall conductivity $\sigma_{xy}^A$. 

We numerically compute the full Nernst coefficient, including contributions from the the conventional $\mathbf{B}$-dependent responses and the anomalous responses, using Eq.~\ref{sxx2}-\ref{ayx2}. In Fig.~\ref{Fig_Nernst_full}, we plot the estimated Nernst coefficient ($\nu/T$) as a function of external magnetic field $\mathbf{B}$ applied in the $\hat{z}$ direction. As suggested by our previous discussion, the Nernst response is primarily dominated by the anomalous contribution. This is further confirmed by our numerical results, where the conventional Nernst coefficient was found to be at least one order of magnitude smaller than the anomalous Nernst coefficient. At lower temperatures ($\sim T<100K$), the behavior of the Nernst coefficient ($\nu/T$) is characterized by an almost step like profile at $\mathbf{B}=0$. The distribution of the flux of the Berry curvature determines the anomalous Hall conductivity $\sigma_{xy}^A$. For $\mathbf{B}<0$, $\sigma_{xy}^A<0$, and for $\mathbf{B}>0$, $\sigma_{xy}^A>0$. Exactly at $\mathbf{B}=0$, one does not expect a finite $\sigma_{xy}^A$, or a finite $\nu$, as $\boldsymbol{\Omega}_{(\mathbf{B}=0)}=0$. As pointed out before, the sign of $\nu$ directly correlates with the sign of $\sigma_{xy}^A$.
\section{Nernst response for inversion breaking WSM}
Our motivation for this section comes from another recent experiment, where the Nernst response in an inversion asymmetric Weyl semimetal has been measured~\cite{Sarah}.
As in a DSM discussed in previous sections, inversion broken WSMs do not break TR, therefore we do not expect anomalous (induced by Berry curvature) Nernst response in the absence of a magnetic field. In the presence of broken TR symmetry by a magnetic field, however, we expect both anomalous and conventional Nernst response.
In this section we calculate the total (sum of conventional and Berry curvature induced) Nernst response for an inversion breaking WSM. The low energy lattice Hamiltonian for an inversion breaking WSM can be written as \cite{McCormickPRB},
\begin{align}
H(\mathbf{k})&=-(m(1-(\cos k_z )^2-\cos k_y )\nonumber \\
&+2t_x(\cos k_x -\cos k_0 ))\sigma_1-2t\sin k_y \sigma_2-2t\cos k_z \sigma_3
\label{Eq_H_inv}
\end{align}
The Hamiltonian in Eq.~(\ref{Eq_H_inv}) produces four Weyl points at $(\pm \pi/2,0,\pm \pi/2)$ as shown in Fig.~6.
We will briefly comment about the symmetries of the above lattice Hamiltonian under the action of two symmetry operators, namely $\mathcal{P}$ (inversion symmetry operator) and $\mathcal{T}$ (time-reversal symmetry operator). Following Ref.~[\onlinecite{McCormickPRB}], we can choose a definite representation of the operators, $\mathcal{P}=\sigma_1$ and $\mathcal{T}=\mathcal{K}$, where $\mathcal{K}$ is the complex conjugation operator (as we are considering spinless fermion bands), we note that $\mathcal{P^\dagger}H(-\mathbf{k})\mathcal{P}\neq H(\mathbf{k})$, and $\mathcal{T}^\dagger H(-\mathbf{k})\mathcal{T}=H(\mathbf{k})$, indicating that the low energy effective Hamiltonian breaks inversion symmetry but preserves time-reversal symmetry.  
To account for contribution from an external magnetic field ($b_z$) we add the term
$b_z \sigma_3$ to the above Hamiltonian.
\begin{figure}[h]
\includegraphics[scale=.3]{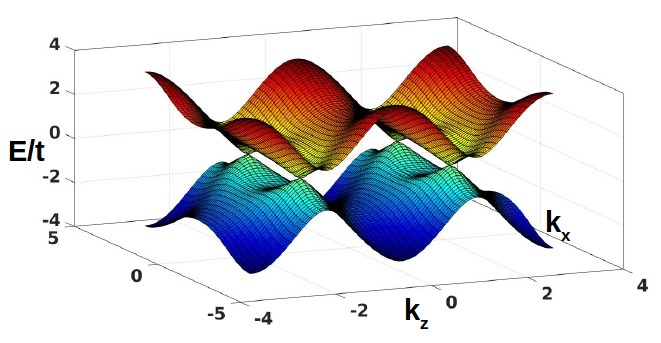}
\includegraphics[scale=0.25]{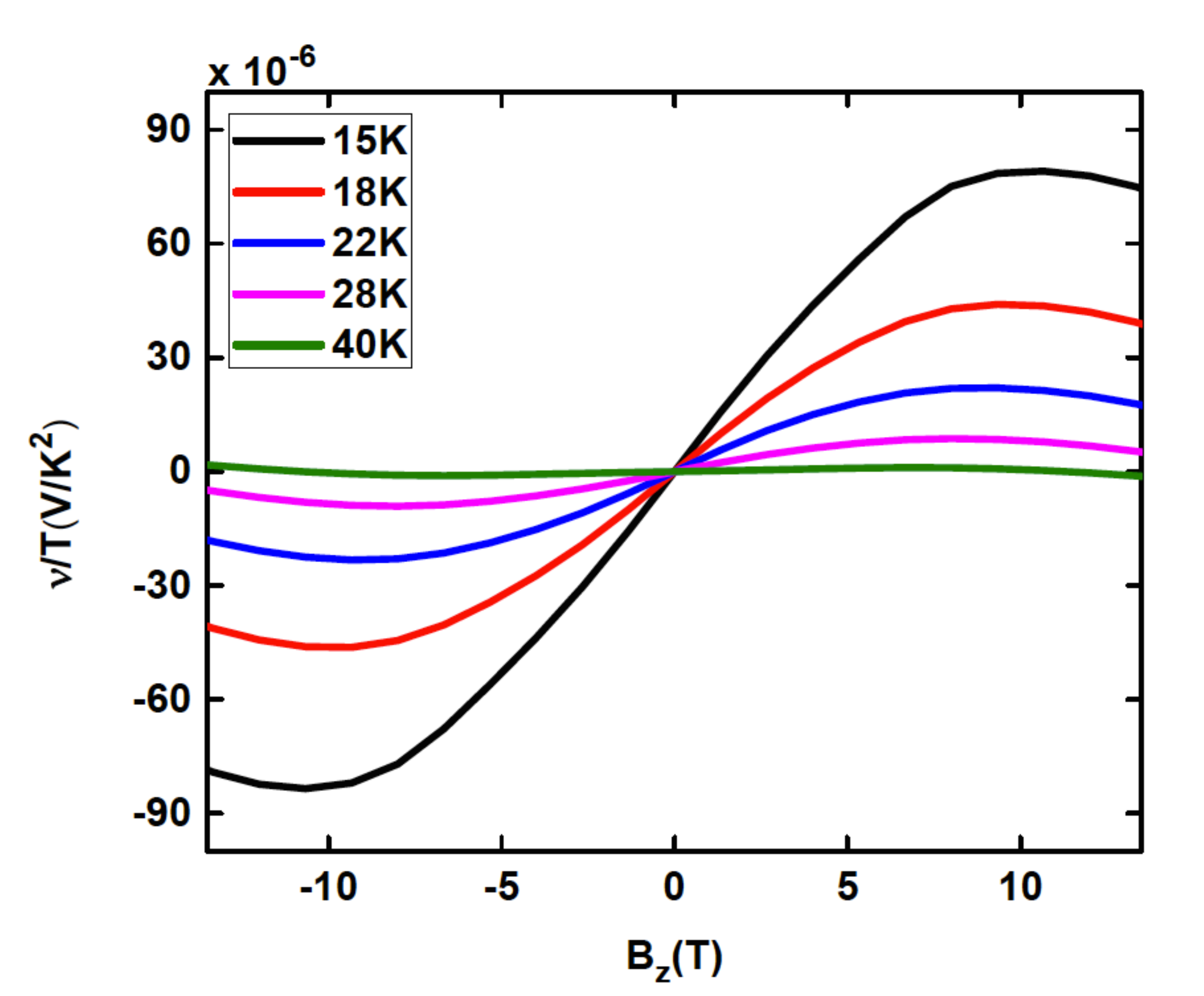}
\caption{ Upper panel: Band structure of inversion broken WSM given by Eq.~(\ref{Eq_H_inv}) consists of four Weyl points located at $(\pm\pi/2,0,\pm\pi/2)$. For our numerical calculations we have used the parameters ~\cite{KlotzPRB,McCormickPRB} $k_0=\pi/2$, $t_x=t/2$, $m=2t$, $\mu=t/5$, $t= 5 meV$, $\tau=0.1 ps$. Lower panel: Total Nernst signal ($ \nu /T$) as a function of applied magnetic field B (in Tesla) for an inversion broken WSM described by the Hamiltonian in Eq.~\ref{Eq_H_inv} at $\mu=t/5$. The plot shows the full Nernst signal for different temperatures. When the magnetic field is zero the total Nernst signal vanishes, as expected, and it gradually decreases with increase in temperature.  Although the full Nernst signal has been plotted, in contrast to the case of DSM (Fig.~\ref{Fig_Nernst_full}), the conventional contribution is significantly larger than the anomalous contribution. As a result, the total Nernst signal has a maximum at a finite value of the magnetic field, in contrast to the Nernst response in DSMs dominated by the anomalous contribution which has a step-like profile near $B=0$. }
\label{Fig:invnernst}
\end{figure}
The total Nernst coefficient normalized by the temperature for inversion broken WSM described by the Hamiltonian in Eq.~(\ref{Eq_H_inv}) is shown in Fig.~\ref{Fig:invnernst}. For this model we find that the conventional contribution is much higher than the anomalous contribution, therefore the Nernst signal constitutes mostly conventional $B-$dependent signal. The Nernst coefficient $\nu/T$ is $\sim 100 \mu V/K^2$, which is of the same order of magnitude as in the experiments \cite{Sarah}. We also note that the Nernst signal has a maximum at a finite value of the magnetic field, in contrast to the behavior of anomalous contribution dominated Nernst coefficient in DSMs which peaks and saturates at low magnetic fields producing a step-like feature near $B=0$ (Fig.~\ref{Fig_Nernst_full}). Our results consisting of a finite magnetic field peak for conventional Nernst signal and the low field saturation leading to a step-like feature near $B=0$ for anomalous Nernst signal are consistent with the experiments \cite{TianLiang:2016}.
\section{Experimental implications}
Recently an anomalous Nernst signal has been reported in thermoelectric experiments on topological Dirac semimetals~\cite{TianLiang:2016}, with a step-like profile similar to our theoretical prediction. In this section we will briefly compare our findings to these recent experiments. The Nernst signals observed in these experiments reveal a large anomalous Nernst coefficient, suggesting the existence of Berry curvature produced by the separation of the Weyl nodes. For samples which reveal an anomalous response, the observed anomalous Nernst signal shows a step-like profile near $B=0$, with significant beating effect in the quantum oscillations of the Nernst signals at higher $B$. The step-like characteristic feature is similar to our prediction, however the beating effect is not revealed in our calculations indicating breakdown of the quasi-classical limit at higher values of $B$. On the other hand, samples which reveal a conventional response, the Nernst signal rises steeply to a sharp Drude-like peak at a finite value of $B$ and then decreases towards zero for higher $B$. This feature is also captured in Fig.~\ref{Fig:invnernst}, which shows the Nernst coefficient for inversion broken WSM which is dominated by the conventional contribution. The order of magnitude of anomalous Nernst signal observed in the experiments also seems to be at least one order of magnitude bigger than the observed conventional Nernst signal, as in our calculations. The orders of magnitude of the Nernst coefficient predicted in our work for Dirac semimetals (Fig.~\ref{Fig_Nernst_full}) and for inversion asymmetric Weyl semimetals (Fig.~\ref{Fig:invnernst}) are similar to the ones seen in experiments in Ref.~[\onlinecite{TianLiang:2016}] and Ref.~[\onlinecite{Sarah}], respectively. In particular in Ref.~[\onlinecite{Sarah}] the Nernst thermopower ($\alpha_{xyz}$)  for $T=10.5K$ is of the order of $100\mu V/K$. On the other hand from our numerical calculations in Fig.~\ref{Fig:invnernst} the Nernst coefficient $\nu/T$ varies between $15\mu V/K^2$ to about $60 \mu V/K^2$ for temperatures between $T=22K$ and $T=15K$ respectively, which translates into a value for the Nernst thermopower ranging from $300\mu V/K-900 \mu V/K$. The slight disagreement between experimental values and numerical ones can be due to uncertainty in the values of $\tau$ which we take to be $\sim 0.1 ps$, which is only an order of magnitude estimate~\cite{KlotzPRB,McCormickPRB}. There is a similar factor of 2-5 difference between the experimentally reported values of $\nu/T$ in Ref.~[\onlinecite{TianLiang:2016}] ($\nu/T\sim 1 \mu V/K^2$)~\cite{TianLiang:2016}, and our numerical values for the same quantity  given in Fig.~\ref{Fig_Nernst_full}, although they are similar in the order of magnitude estimates. This slight discrepancy can also be due to uncertainties in the values of $\tau$ which again is taken to be $\tau\sim 0.1ps$ as in Fig.~\ref{Fig:invnernst}. 
\section{Conclusions}
Three dimensional topological Dirac semimetals are characterized by nodes in the bulk energy spectrum with a vanishing Chern number. Because of the vanishing flux of the Berry curvature through any surface in the Brillouin zone, anomalous Hall and Nernst conductivities vanish in a topological DSM in the absence of a magnetic field. In the presence of a magnetic field, however, a topological DSM reduces to a Weyl semimetal, evincing a non-zero Hall and Nernst response, which have contributions from both conventional as well as anomalous (Berry curvature mediated) components. In this paper we consider a topological DSM with a pair of Dirac nodes on a high symmetry axis (axis of four fold rotational symmetry, $C_4$), which is an appropriate description of the experimentally realized Dirac semimetal Cd$_3$As$_2$. For this system we compute the total Nernst coefficient (conventional as well as anomalous) in the presence of a small finite magnetic field (~a few Tesla) and small chemical potential, within the Boltzmann description in the relaxation time approximation. 

In Boltzmann formalism we find that the conventional Nernst response in topological DSMs is typically small due to Sondheimer cancellation, and can be of either sign depending on temperature, small and negative at low temperatures ($\sim 10$ K), and small and positive at higher temperatures ($\sim 100$ K). In contrast, we find that the anomalous Nernst response is large and positive because of the peaking of the Berry curvature in the limit of small chemical potentials and due to the absence of Sondheimer cancellation. Our calculated anomalous Nernst coefficient is almost one order of magnitude larger than the conventional Nernst coefficient at similar temperatures and magnetic field. The measured low-field Nernst coefficient, thus, is expected to be dominated by the anomalous Nernst effect, at least in the limit of small temperatures. At low temperatures, the behavior of the total Nernst coefficient is characterized by an almost step like profile at $\mathbf{B}=0$.  However, exactly at $\mathbf{B}=0$ there is no Nernst signal, from either conventional or anomalous contributions, because of the restoration of time reversal symmetry.  Our results have direct experimental relevance for Nernst and thermoelectric measurements on the experimentally available topological DSMs Cd$_3$As$_2$ and Na$_3$Bi, as an anomalous Nernst signal can directly probe Berry curvature in these materials. Recently an anomalous Nernst signal has been reported in thermoelectric experiments on topological Dirac semimetals~\cite{TianLiang:2016}, with a step-like profile similar to our theoretical prediction.

Additionally we have also discussed Nernst response of an inversion asymmetric WSM. Since both the DSM and the inversion broken WSM are TR invariant systems, the Nernst effect in inversion broken WSM is thus expected to be similar to that in a DSM. However, we find that in the inversion broken WSM the conventional Nernst coefficient dominates over the anomalous contribution which is opposite to the case in DSM. Our calculation on Nernst response in an inversion broken WSM is also directly relevant to recent experiments in NbP~\cite{Sarah}.

We acknowledge support from ARO Grant No: (W911NF-16-1-0182).


\end{document}